# Universal method of strictly calculating self-consistent fields of realistic plasma particles

H. Lin

*State Key Laboratory of High Field Laser Physics, Shanghai Institute of*

*Optics and Fine Mechanics,*

*P. O. Box 800-211, Shanghai 201800, China;*

*linhai@siom.ac.cn*

()

A universal method of strictly calculating self-consistent fields of realistic plasma particles could be strictly derived from three basic tools in theoretical plasma physics: particle simulation, Vlasov-Maxwell theory and fluid theory.

Plasma physics is a physical branch about many charged particles interacting through their self-consistent fields. In its earlier developing stage (about 1940s~1960s), many theoretical methods [1-6] which are successful in other elder physical branch such as neutral gas physics and fluid mechanics were transplanted into this younger branch and rapidly built up the basis of this new branch. However, almost no one doubts whether these transplanted methods are appropriate for plasmas where numerous charged particles are correlated through their self-consistent fields. More important, in above-mentioned transplanted methods the plasma self-consistent fields is never strictly calculated but is indeed treated by various (obvious and hidden) approximations.

Although people have realized the importance of strictly calculating plasma self-consistent fields, some realistic adverse factors prevents this goal being thoroughly achieved. Let us comment three basic tools in plasma physics one by one.

a) particle simulation. [7-9]

This basic tool is to solve $2N+4$ equations describing $N$ realistic particles (or macroparticles)

$$\partial_t E(R,t) = \nabla \times B(R,t) + \sum_i d_t r_i(t) \delta(r_i(t) - R); \qquad (P.1)$$

$$\partial_t B(R,t) = -\nabla \times E(R,t); \qquad (P.2)$$

$$\nabla \cdot E(R,t) = \sum_i \delta(r_i(t) - R); \qquad (P.3)$$

$$\nabla \cdot B(R,t) = 0; \qquad (P.4)$$

...

$$d_t \frac{d_t r_i(t)}{\sqrt{1 - [d_t r_i(t)]^2}} = E(r_i(t), t) + d_t r_i(t) \times B(r_i(t), t); \qquad (P.2i+4)$$

$$v_i = d_t r_i. \qquad (P.2i+5)$$

...

It is in principle a strict method. But in practice, because $N$ is nearly an astronomical figure, people often resort to an approximation method, which is often called Particle-In-Cell (PIC) method. The PIC method is to approximate the solution of above $2N + 4$ equations with that of $2N/R_{merge} + 4$ equations, where $R_{merge} > 1$ measure how many realistic particles are merged into a so-called macroparticle, and to alternatively updating $N/R_{merge}$ macroparticles' information (i.e., position and velocity) and $(E, B)$.

b) Vlasov-Maxwell theory. [1-6]

This basic tool is to solve 5 equations

$$\partial_t E(R,t) = \nabla \times B(R,t) + \int v f d^3 p; \qquad (V.1)$$

$$\partial_t B(R,t) = -\nabla \times E(R,t); \qquad (V.2)$$

$$\nabla \cdot E(R,t) = \int f d^3 p - N_i; \qquad (V.3)$$

$$\nabla \cdot B(R,t) = 0; \qquad (V.4)$$

$$\partial_t f + v \cdot \nabla f - [E + v \times B] \cdot \partial_p f = 0. \qquad (V.5)$$

It is also in principle a strict method. But in practice, because $f$ is defined over a 6-D phase space and hence corresponds to too huge data mount, if above 5 equations are solved by alternatively updating $f$ and $(E, B)$, updating $f$ will be very time-consuming. This basic tool is therefore less applied than the approximation version of particle simulation, i.e., PIC method.

c ) fluid theory. [1-6]

This basic tool is to solve $5+1$ equations

$$\partial_t E(R,t) = \nabla \times B(R,t) + nu_{fl}; \tag{F}$$

$$\partial_t B(R,t) = -\nabla \times E(R,t); \tag{F}$$

$$\nabla \cdot E(R,t) = n - N_i; \tag{F}$$

$$\nabla \cdot B(R,t) = 0; \tag{F}$$

$$\partial_t p_{fl} + u_{fl} \cdot \nabla p_{fl} = E + u_{fl} \times B + \text{thermal pressure/density}; \tag{F}$$

$$\text{Assumed thermodynamical state equation (about thermal pressure/density)} \tag{F}$$

Because of assumed thermodynamical state equation, this basic tool is often viewed as inferior than other two tools. The self-consistent fields $(E, B)$ obtained from this tool is hence taken as less reliable than its counterparts obtained from other tools.

Following text will show in details that if above three basic tools are in their respective strict forms, they will agree with each other to yield a strict method of calculating $(E, B)$ of realistic particles.

a) for particle simulation

We could rewrite any relativistic Newton equation, for example Eq.(P.2i+4), as

$$\begin{aligned} 0 = & d_t \frac{d_t r_i(t)}{\sqrt{1 - [d_t r_i(t)]^2}} - E(r_i(t), t) - d_t r_i(t) \times B(r_i(t), t) \\ = & \left[ \left[ d_t \frac{d_t r_i(t)}{\sqrt{1 - [d_t r_i(t)]^2}} - d_t \frac{u(r_i(t), t)}{\sqrt{1 - [u(r_i(t), t)]^2}} \right] - [d_t r_i(t) - u(r_i(t), t)] \times B(r_i(t), t) \right. \\ & \left. + \left[ d_t \frac{u(r_i(t), t)}{\sqrt{1 - [u(r_i(t), t)]^2}} - E(r_i(t), t) - u(r_i(t), t) \times B(r_i(t), t) \right], \right. \end{aligned} \tag{fp.1}$$

and note a fact that it is valid for arbitrary value of $d_t r_i(t)$, or arbitrary value of $\Delta = d_t r_i(t) - u(r_i(t), t)$. Eq.(fp.1) is of a binary-function type general form: $0 = function(var1, var2)$, where $var1$ and $var2$ are independent variables. Thus, timing Dirac function $\delta(\Delta)$ at both side of Eq.(fp.1) and then integrating over $\Delta$, we could obtain

$$0 = \int [\text{right-side terms of Eq.(fp.1)}] * \delta(\Delta) \, d\Delta$$

$$= \left[ d_t \frac{u(r_i(t), t)}{\sqrt{1 - [u(r_i(t), t)]^2}} - E(r_i(t), t) - u(r_i(t), t) \times B(r_i(t), t) \right]. \qquad \text{(fp.2.b)}$$

Here, we have utilized, when deducing Eq.(fp.2.b), following property of the Dirac function: $x\delta(x) = 0$. This property immediately leads to $[d_t x] * \delta(x) = -x * [d_t \delta(x)]$. Noting the property of $d_t \delta(x)$: $d_t \delta(x) = 0$ if $d_t x = 0$; and $d_t \delta(x) = \delta(x)$ if $d_t x \neq 0$, (i.e., if $x$ varies with respect to $t$, $x$-value will derivate from 0 and corresponding $\delta(x)$-value will also jump from $\infty$ to 0), we could find that there are $x * [d_t \delta(x)] = 0$ if $d_t x = 0$ and $x * [d_t \delta(x)] = x\delta(x) = 0$ if $d_t x \neq 0$, i.e, no matter what $d_t x$-value is, there is always $x * [d_t \delta(x)] = 0$, and hence $[d_t x] * \delta(x) = 0$. The integral $\int [\text{right-side terms of Eq.(fp.1)}] * \delta(\Delta) \, d\Delta$ includes terms of a general form $\int d_t \Delta * \delta(\Delta) \, d\Delta$. These properties of the Dirac function lead to $\int d_t \Delta * \delta(\Delta) \, d\Delta = 0$.

Subtracting Eq.(fp.1) and Eq.(fp.2.b), we have

$$0 = \left[ d_t \frac{d_t r_i(t)}{\sqrt{1 - [d_t r_i(t)]^2}} - d_t \frac{u(r_i(t), t)}{\sqrt{1 - [u(r_i(t), t)]^2}} \right] - [d_t r_i(t) - u(r_i(t), t)] \times B(r_i(t), t).$$

(fp.2.a)

Therefore, any solution of Eqs.(P) is also that of following equation set of $2N + 5$ members

$$0 = \left[ d_t \frac{u_{fl}(r_i(t), t)}{\sqrt{1 - [u_{fl}(r_i(t), t)]^2}} - E(r_i(t), t) - u_{fl}(r_i(t), t) \times B(r_i(t), t) \right] \qquad (2.0)$$

$$\partial_t E(R, t) = \nabla \times B(R, t) + \sum_i d_t r_i(t) \delta(r_i(t) - R); \qquad (2.1)$$

$$\partial_t B(R, t) = -\nabla \times E(R, t); \qquad (2.2)$$

$$\nabla \cdot E(R,t) = \sum_i \delta\left(r_i(t) - R\right); \tag{2.3}$$

$$\nabla \cdot B(R,t) = 0; \tag{2.4}$$

...

$$\left[d_t \frac{d_t r_i(t)}{\sqrt{1 - [d_t r_i(t)]^2}} - d_t \frac{u_{fl}(r_i(t),t)}{\sqrt{1 - [u_{fl}(r_i(t),t)]^2}}\right] = [d_t r_i(t) - u_{fl}(r_i(t),t)] \times B(r_i(t),t) \tag{2.2i+4}$$

$$v_i = d_t r_i \tag{2.2i+5}$$

...,

where $u_{fl}(R,t) = \sum_{i \in r_i(t)=R} d_t r_i(t) / \sum_{i \in r_i(t)=R} 1$. On the other hand, it is obvious that any solution of Eqs.(2) is also be that of Eqs.(P). In a mathematical language, Eqs.(P) and Eqs(2) have their respective solution sets: {solutions of Eqs.(2)} and {solutions of Eqs.(P)}, and there strictly exists a relation between these two sets: {solutions of Eqs.(2)} = {solutions of Eqs.(P)}. Namely, starting from the starting model equations of particle simulation scheme, we could find that there exists a closed equation set of $u_{fl}$, $E$ and $B$, i.e., Eqs.(2.0-4).

b) for Vlasov-Maxwell theory.

We could rewrite Vlasov equation (VE), for example Eq.(V.5), as

$$\begin{aligned}
0 &= \partial_t f + v \cdot \nabla f - [E + v \times B] \cdot \partial_p f. \\
&= [\partial_t (f - f_{mono}) + v \cdot \nabla (f - f_{mono}) - [E + v \times B] \cdot \partial_p (f - f_{mono})] \\
&+ (v - u_{fl}) \cdot \nabla f_{mono} - (v - u_{fl}) \cdot \partial_p f_{mono} \\
&+ [\partial_t f_{mono} + u_{fl} \cdot \nabla f_{mono} - [E + u_{fl} \times B] \cdot \partial_p f_{mono}]. \tag{fV.1}
\end{aligned}$$

Any distribution function $f$ has two independent characteristic parameters: the variance and the mean. Here, the mean of $f$ is represented by $u_{fl} = \int v f d^3 p / \int f d^3 p$). For any distribution $f$, we could express it as $f = n\delta(v - u_{fl}) + a_0 \delta(v - u_{fl}) + \sum_{i \geq 1} a_i (v - u_{fl})^i$ (where $n = \int f d^3 p$, $u_{fl} = \int v f d^3 p / \int f d^3 p$, $a_i$ are independent of $v$, $a_0$ depends on all coefficients $a_{i \geq 1}$ through two relations, $\int \left[a_0 \delta(v - u_{fl}) + \sum_{i \geq 1} a_i (v - u_{fl})^i\right] d^3 p = 0$ and $\int v \left[a_0 \delta(v - u_{fl}) + \sum_{i \geq 1} a_i (v - u_{fl})^i\right] d^3 p = 0$, i.e. $a_0$ is a function of all $a_{i \geq 1}$, $a_0 = a_0(a_1, ..., a_i, ...)$. Here, a given pair of $(n, u_{fl})$ could correspond to multiple possible distribution modes over $v$-space. This fact determines these two relations). Substituting this expression into VE and comparing the coefficients of $(v - u_{fl})^i$-term, we could find that there exists following equation for $f_{mono} = n\delta(v - u_{fl}) + a_0 \delta(v - u_{fl})$ (because of the fact that VE is valid at any $v$-value.)

$$0 = \partial_t f_{mono} + u_{fl} \cdot \nabla f_{mono} - [E + u_{fl} \times B] \cdot \partial_p f_{mono}, \qquad \text{(fV.2)}$$

which could directly lead to (here, as stressed latter, $p(u_{fl}) = \frac{u_{fl}}{\sqrt{1 - u_{fl}^2}}$)

$$0 = \partial_t [p(u_{fl})] + u_{fl} \cdot \nabla_r [p(u_{fl})] - [E + u_{fl} \times B] \qquad \text{(fV.3)}$$

according to standard procedure, i.e., two relations, $\int [\text{right-side terms in Eq.(fV.2)}] d^3 p = 0$ and $\int p * [\text{right-side terms in Eq.(fV.2)}] d^3 p = 0$, will lead to Eq.(fV.3), which is equivalent to Eq.(2.0)

Indeed, because Eq.(fV.1) is of a binary-function type general form: $0 = function(var1, var2)$, where $var1$ and $var2$ are independent variables, like deriving Eq.(fp.2.b) from Eq.(fp.1), we could derive Eq.(fV.2) similarly (where $var2 = v - u_{fl}$)

$$0 = \int [\text{right-side terms of Eq.(fV.1)}] * \delta(var2) \, dvar2$$
$$= \text{right-side terms in Eq.(fV.2)}. \qquad \text{(fV.2)}$$

c) for fluid theory

It is well-known that fluid theory is indeed a derivant of V-M theory. All equations in fluid theory, except the assumed thermodynamical state equation, could be derived from 5 equations in V-M theory according to standard procedure. For example, two relations, $\int [\text{right-side terms in Eq.(fV.1)}] d^3 p = 0$ and $\int p * [\text{right-side terms in Eq.(fV.1)}] d^3 p = 0$, will lead to Eq.(F.5).

Because the velocity $v$ is a nonlinear function of the momentum $p$ (i.e., $v = \frac{p}{\sqrt{1 + p^2}}$) and vice versa, we should note that the statistic average value $\int p f d^3 p / \int f d^3 p$ (i.e. fluid momentum $p_{fl}$) is usually not equal to the momentum corresponded by the statistic average value $\int vfd^3p/\int fd^3p$ (or fluid velocity $u_{fl}$), i.e., $p_{fl} \neq p(u_{fl})$(where $p(u_{fl})$ refers to the value of function $p(variable)$ at $variable = u_{fl}$), if the distribution $f$ is not a Dirac function of $p$ (i.e., $f$ has a thermal spread over $p$-space). Only at zero temperature case, there is $p_{fl} = p(u_{fl})$. (Strictly speaking, if $f$ is a symmetric function of $p$, there will be $p_{fl} = p(u_{fl}) = 0$, $u_{fl} = 0$ and thermal pressure$\neq 0$. But this special case corresponds to $E = 0$ and $E + u_{fl} \times B = 0$. A non-zero thermal pressure will drive $p_{fl}$ differing from 0 according to Eq.(F.5). Once $p_{fl} \neq 0$, there will be $p_{fl} \neq p(u_{fl})$ because $f$ has an asymmetric thermal spread over $p$-space).

After noting the difference between $p_{fl}$ and $p(u_{fl})$, a scrupulous reader will also note that it is $u_{fl}$, rather than $p_{fl}$, that appears in Maxwell equations (*Meqs*). More important, he might consider whether or not there is necessity to introduce an assumed thermodynamical state equation. This is because two relations, $\int[\text{right-side terms in Eq.(fV.2)}]d^3p = 0$ and $\int p * [\text{right-side terms in Eq.(fV.2)}]d^3p = 0$, will lead to Eq.(fV.3). Substracting Eq.(F.5) and Eq.(fV.3), an equation about thermal pressure/density will naturally appear.

Therefore, we could strictly derive fluid theory from V-M theory according to a standard procedure without introducing any assumption. This makes fluid theory becoming really basic tool whose reliability is equal to those of other tools. In other words, the assumed thermodynamical state equation, Eq.(F.6), is replaced by $Eq.(F.5) - Eq.(fV.3)$. Moreover, even if starting from particle simulation, we could still find that for all subindex $i$ meeting $i \in r_i(t) = R$, summing corresponding Eq.(fp.2.a) will also lead to $Eq.(F.5) - Eq.(fV.3)$. This also suggests that particle simulation and Vlasov-Maxwell theory completely agree with each other.

By now, we have displayed in details how to obtain a closed equation set of $u_{fl}$, $E$ and $B$ from three basic tools. In short, no matter which one of three basic tools is chosen by people when investigating plasma physics, $E$ and $B$, obey a fixed fluid equation set, Eqs.(2.0-4). Indeed, these different basic methods are equivalent if they are in their respective strict forms. There is no reason to think that any method is better than others.